# Scalable native signed optical computing enabled by dual-wavelength incoherent multiplexing

Yuan Ren[1,2], Yong Zheng[2,*], Ruixue Liu[1,2], Yunpeng Song[2], Qinfen Huang[1,2], Min Wang[2,*] and Ya Cheng[1,2,3,4,5,*]

[1]State Key Laboratory of Precision Spectroscopy, East China Normal University, Shanghai 200062, China.
[2]The Extreme Optoelectromechanics Laboratory (XXL), School of Physics, East China Normal University, Shanghai 200241, China.
[3]Hefei National Laboratory, Hefei 230088, China.
[4]Shanghai Research Center for Quantum Sciences, Shanghai 201315, China.
[5]Collaborative Innovation Center of Extreme Optics, Shanxi University, Taiyuan 030006, China.
*Correspondence: Yong Zheng (yzheng@phy.ecnu.edu.cn), Min Wang (mwang@phy.ecnu.edu.cn), Ya Cheng (ya.cheng@siom.ac.cn).

## ABSTRACT

Incoherent photonic neural networks (PNNs) provide a robust platform for analog optical computing, yet efficient implementation of native signed operations remains challenging. Existing incoherent PNNs approaches often require additional spatial channels or temporal encoding steps to represent bipolar input signals, resulting in hardware overhead that scales with system size. Here, we demonstrate a dual-wavelength incoherent photonic architecture that natively supports both signed inputs and signed weights on a thin-film lithium niobate platform. By encoding complementary signal components onto two wavelength channels and performing computation within a shared physical path, the proposed scheme eliminates duplicated weighting units. As a result, the additional hardware overhead associated with signed computation remains constant per multiply–accumulate operation, independent of matrix size. The fabricated device exhibits a modulation bandwidth exceeding 40 GHz and achieves four-quadrant optical multiplication with a standard deviation error of 1.27%. System-level functionality is validated through neural-network classification, achieving 95.1% accuracy on the Moons dataset and 91.63% on MNIST. These results establish a practical route toward scalable incoherent photonic computing systems with native bipolar processing capability.

## INTRODUCTION

The rapid growth of artificial intelligence (AI) workloads has driven increasing demand for computing hardware capable of overcoming the energy-efficiency and throughput limitations of conventional von Neumann architectures.[1-6] Physical neural networks which directly exploit physical processes to perform analog computation, have therefore emerged as a promising route toward next-generation intelligent computing systems.[7,8] Among various physical implementations, optical neural networks (ONNs) have attracted extensive interest owing to the intrinsic advantages of photonic carriers, including ultrahigh bandwidth, low latency, and massive parallelism.[9-11] These characteristics enable optical systems to potentially achieve computational throughput and energy efficiency beyond that of electronic accelerators, particularly for the large-scale matrix operations that dominate modern deep learning workloads. Coherent optical neural networks based on Mach–Zehnder interferometer (MZI) meshes have been widely investigated as a representative programmable photonic computing architecture.[9,12-24] By controlling the optical phase and interference within cascaded MZI networks, these systems can implement arbitrary unitary matrix transformations and therefore provide a universal framework for optical linear computing.[12-14] However, as the network dimensionality increases, coherent MZI meshes suffer from accumulated optical loss, phase instability, calibration complexity, and fabrication-induced errors. These issues substantially degrade computational fidelity and significantly limit the scalability of large scale coherent photonic processors.

To address the scalability limitations of coherent architectures, increasing attention has recently been directed toward incoherent optical computing schemes.[10,11,25] Incoherent optical neural networks eliminate the stringent phase-stabilization requirements associated with coherent interference and therefore exhibit improved robustness, simpler calibration procedures, and favorable scalability. Nevertheless, because optical intensity is intrinsically non-negative, most demonstrated incoherent multiply-accumulate (MAC) systems remain fundamentally restricted to non-negative signal representations. Although balanced detection can partially support signed weights through differential readout, the input data themselves are typically constrained to positive-valued encoding.[11,25] This limitation is particularly problematic for practical machine learning applications, where zero-centered bipolar data distributions are essential for efficient optimization and training convergence.[2,5,6,26-28] Several approaches have recently been explored to support signed computation in photonic systems, including spatially separated differential encoding,[29] temporal multiplexing strategies,[25] and nonlinear transfer-function encoding schemes based on trigonometric modulation responses.[23] However, these methods generally require either proportional hardware duplication, additional computing cycles, or complex system-level compensation, thereby reducing integration density and limiting scalability for large-dimensional tensor operations.

Here, we propose a scalable dual-wavelength incoherent computing architecture that natively supports both signed inputs and signed weights on a thin-film lithium niobate (TFLN) platform. By encoding complementary signal components onto two spectrally separated optical carriers, the proposed scheme realizes true four-quadrant multiplication within a shared wavelength-division multiplexed computing structure. Meanwhile, the proposed architecture introduces only a fixed spectral encoding overhead for each signed computing channel as the system scales. Experimentally, we demonstrate a high-speed signed optical multiplier with the electro-optic bandwidth exceeding 40 GHz and validate its computational capability through nonlinear classification and handwritten digit recognition tasks. Combined with the wafer-scale compatibility of the photolithography-assisted chemo-mechanical etching process, this work establishes a compact and scalable hardware foundation for future large-scale incoherent photonic neural networks.

## ARCHITECTURE AND OPERATING PRINCIPLE

To illustrate the necessity of native signed computation involving signed inputs and signed weights in photonic neural networks, we first investigate a nonlinear classification task through numerical backpropagation simulations. As shown in Fig. 1(a), neural networks operating with bipolar inputs rapidly converge toward high classification accuracy. In contrast, constraining the input signals to a purely non-negative domain significantly slows the training process and limits the final achievable accuracy, particularly for shallow network architectures. This performance degradation originates from the loss of zero-centered feature distributions, which are critical for efficient gradient propagation and parameter optimization in modern machine learning models.[2,5,6,26-28] These results highlight the importance of implementing native four-quadrant operations directly in optical hardware.

Our incoherent optical multiplier employs a dual-wavelength differential design to natively support both bipolar inputs and bipolar weights. As illustrated in Fig. 1(b), a temporal sequence of bipolar input values $X_i \in [-1,1]^N$ is converted into complementary non-negative signals $x_i^+$ and $x_i^-$, and satisfying

$$X_i = x_i^+ - x_i^- \tag{1}$$

these signals are encoded onto two distinct light beams with wavelengths of $\lambda_1$ and $\lambda_2$ via separate high-speed electro-optic modulators biased near their quadrature points. The two light beams are then injected into an unbalanced MZI-based wavelength multiplexing. The arm length difference $\Delta L$ of the multiplexing is carefully designed to ensure that light beams with wavelengths of $\lambda_1$ and $\lambda_2$ from different input ports are routed to the same output port. The multiplexed signals are then delivered to a shared weighting unit implemented using another unbalanced MZI with an identical spectral response. Consequently, the effective transmission weights $w_i^+$ and $w_i^-$ for $\lambda_1$ and $\lambda_2$ exhibit complementary responses, i.e.,

$$w_i^+ + w_i^- = 1 \tag{2}$$

By taking the difference of the optical intensities at the complementary output ports of each weighting MZI, the final output becomes

$$(w_i^+ - w_i^-)(x_i^+ - x_i^-) = W_i X_i \tag{3}$$

where $W_i = w_i^+ - w_i^- \in [-1,1]^N$. As a result, the proposed architecture directly realizes native four-quadrant optical multiplication within an incoherent intensity-computing framework while maintaining a compact and scalable shared-path computing structure for signed operations.

In addition to enabling native signed computation, the incoherent intensity-based implementation adopted in this work provides several practical advantages over coherent photonic computing schemes. Since the proposed architecture relies on optical intensity rather than phase information, it is inherently insensitive to optical phase noise and environmental perturbations such as temperature fluctuations and fabrication-induced phase errors. This significantly relaxes the requirement for active phase stabilization, which is typically necessary in coherent interferometric computing systems. Furthermore, the use of wavelength-domain encoding allows independent control of different signal channels without introducing coherent crosstalk. This property is particularly beneficial for large-scale integration, where maintaining global phase coherence across complex photonic circuits becomes increasingly challenging.

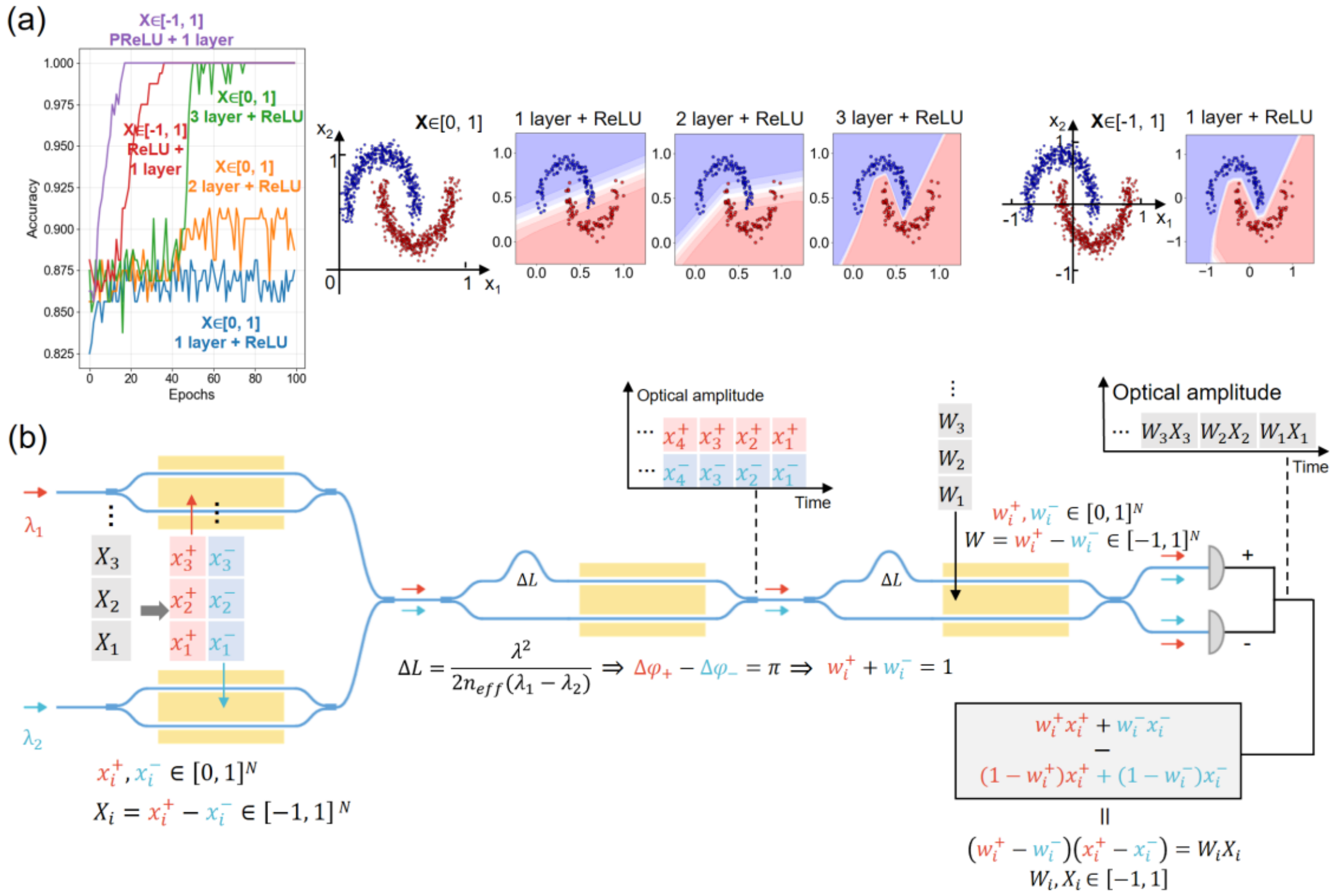

*Fig. 1. Principle of the proposed native signed incoherent optical computing architecture. (a) Simulated nonlinear classification accuracy for bipolar and non-negative neural-network inputs. (b) Schematic of the dual-wavelength differential optical multiplier enabling native four-quadrant optical multiplication.*

To implement the proposed architecture, an initialization and calibration procedure is first performed to determine the corresponding input wavelengths $\lambda_1$ and $\lambda_2$. Specifically, the laser with wavelength $\lambda_1$ = 1550 nm is injected into the data Mach-Zehnder interferometer (MZI), while the two unbalanced MZIs are tuned so that the laser light exits entirely from the + port of the weighting MZI. The 1550 nm laser is then turned off while the state of the weighting MZI remains unchanged. Then the wavelength of a tunable laser injected into the other data MZI is swept until the laser exits exclusively from the - port of the weighting MZI, at which point the second wavelength $\lambda_2$ is obtained, thereby establishing the complementary spectral response required for signed computation. To further suppress the residual imbalance in two-wavelength balanced photodetection, a low-frequency triangular-wave modulation is applied to the weighting MZI while monitoring the transmission responses of both wavelength channels. By fine-tuning the intermediate wavelength-multiplexing MZI, the peak transmission intensities of the two wavelength channels at the complementary output ports are equalized, thereby minimizing offset errors during differential detection. During dynamic operation, radio-frequency signals from an arbitrary waveform generator (AWG) drive the electro-optic modulators for high-speed data encoding.

# RESULTS

## Device fabrication and characterization

The proposed native signed optical multiplier was fabricated on an x-cut TFLN platform using the photolithography-assisted chemo-mechanical etching (PLACE) process.[30-33] Fig. 2(a) presents the optical microscope image of the fabricated chip, which integrates four high-speed electro-optic modulation units forming the dual-wavelength signed-computing architecture within a compact footprint. Further, numerous passive optical elements such as 1×2 multimode interference (MMI) couplers, spot size converter (SSC) and bend waveguides are fabricated using the PLACE technology with high manufacturing uniformity, and smooth waveguide sidewalls over the entire circuit lead to a low optical propagation loss of 0.03 dB/cm. Low optical propagation loss provided by the PLACE process is particularly beneficial for large-scale photonic neural network integration.

To characterize the static electro-optical (EO) property of the fabricated modulators, a 100 MHz triangle wave signal with a peak-to-peak voltage of 20 V was applied on the modulator. The EO responses of the 7-mm-long modulator is shown in Fig. 2(c), from which the $V_\pi$ was measured to be 3.215 V. Thus the extracted $V_\pi \cdot L$ can be calculated as 2.25 V·cm. Subsequently, we characterized the broadband EO response of the fabricated modulators integrated with a $50\ \Omega$ load. The measured electro-optic transmission and electric-electric input reflection as a function of the applied RF frequency from 10 MHz to 40 GHz are shown in Fig. 2(b). It is observed that the modulator achieves a 3-dB bandwidth exceeding 40 GHz while maintaining microwave reflection suppression of more than 20 dB across the measured range.

To experimentally validate the dual-wavelength signed computing mechanism, the wavelength dependent transmission characteristics of weighting MZI are further characterized. Fig. 2(d) shows the normalized transmission spectra measured at the selected operating wavelengths of 1550.0 nm and 1548.4 nm. Both wavelength channels exhibit high extinction ratios exceeding 30 dB. More importantly, as shown in Fig. 2(e), the transmission responses of the shared weighting MZI for the two wavelength channels exhibit complementary modulation behavior with an effective $\pi$-phase spectral offset. The result provides experimental validation of the complementary weighting condition for native signed optical computing as derived in Equation 2.

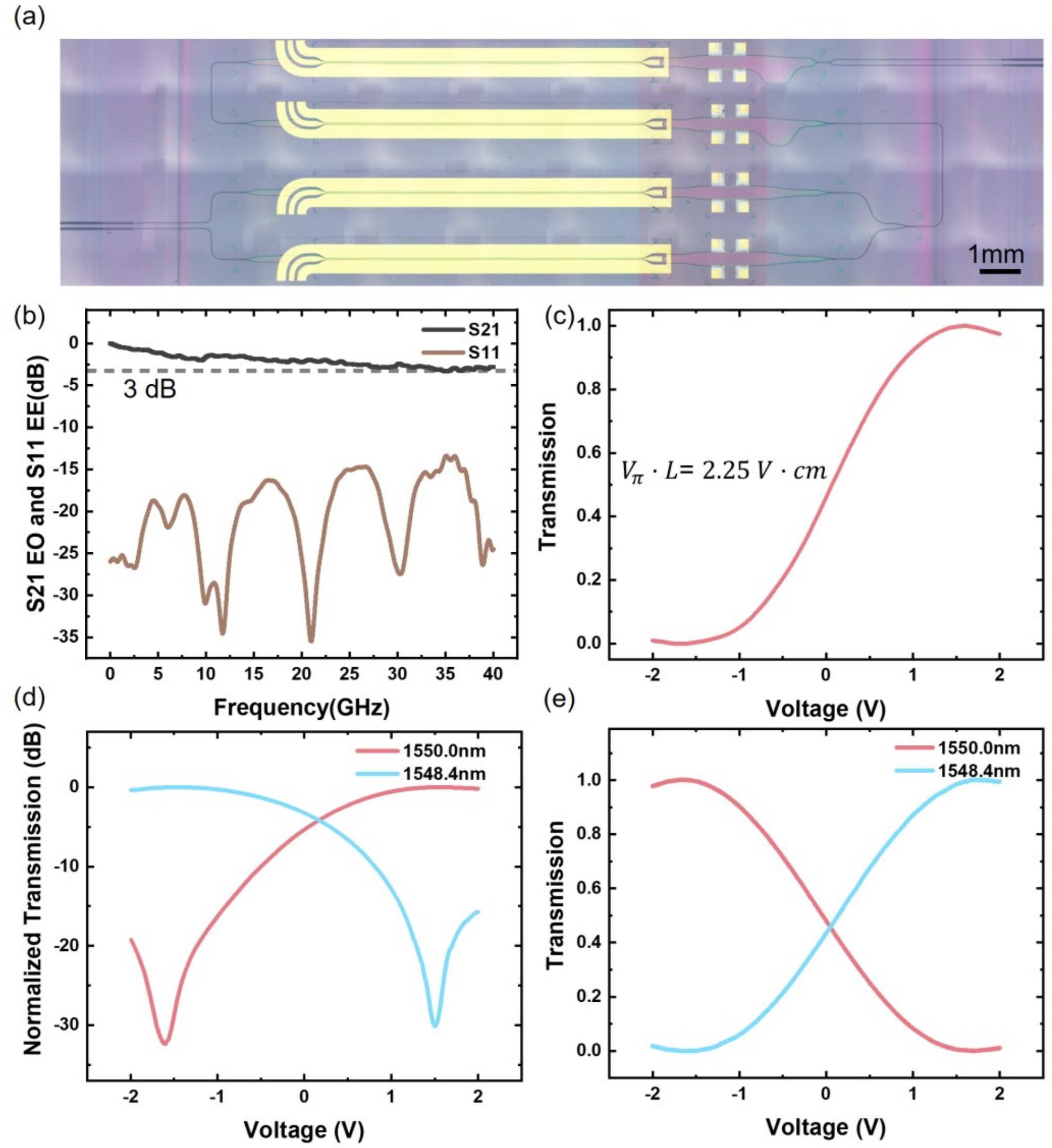

*Fig. 2. Experimental characterization of the fabricated dual-wavelength signed optical multiplier. (a) Optical microscope image of the fabricated TFLN photonic chip. All fabricated modulators are integrated with a 50 Ω load. (b) Measured electro-optic transmission* $(S_{21})$ *and electric-electric input reflection* $(S_{11})$ *response of the 7-mm-long modulator in our circuit. (c) Static transmission response of the electro-optic modulator. (d) Measured transmission spectra at 1550.0 nm and 1548.4 nm. (e) Complementary transmission responses of the shared weighting MZI for the two wavelength channels.*

## Dynamic signed multiplication characterization

To experimentally evaluate the dynamic computing fidelity of the proposed architecture, continuous four quadrant optical multiplication was performed in the time domain. In this experiment, 1000 random bipolar input values $X_i$ and weight values $W_i$ were generated with uniform distributions spanning the range from $-1$ to 1. These vectors are encoded into the time domain of each channel and multiplied after the light sequentially passes through the data and weight modulators. The vectors are first converted from digital to analog signals by a high-speed arbitrary waveform generator. A dual-channel fiber array collects the resulting optically modulated signal, which is then directed to the high-speed photodetector. The test data are subsequently displayed on a high-speed oscilloscope. The computational accuracy per channel can be assessed by comparing the expected values with the photonic dot products, which are obtained from direct electronic integration of the sampled time traces. Figs. 3(a) and 3(b) respectively compare the theoretical target signals and experimentally measured signals for the encoded input data and weighting signals within a $10\,\mu s$ temporal window. The experimentally measured amplitudes closely follow the ideal target distributions, yielding Gaussian-like error histograms with standard deviations of 1.44% and 1.49%, respectively.

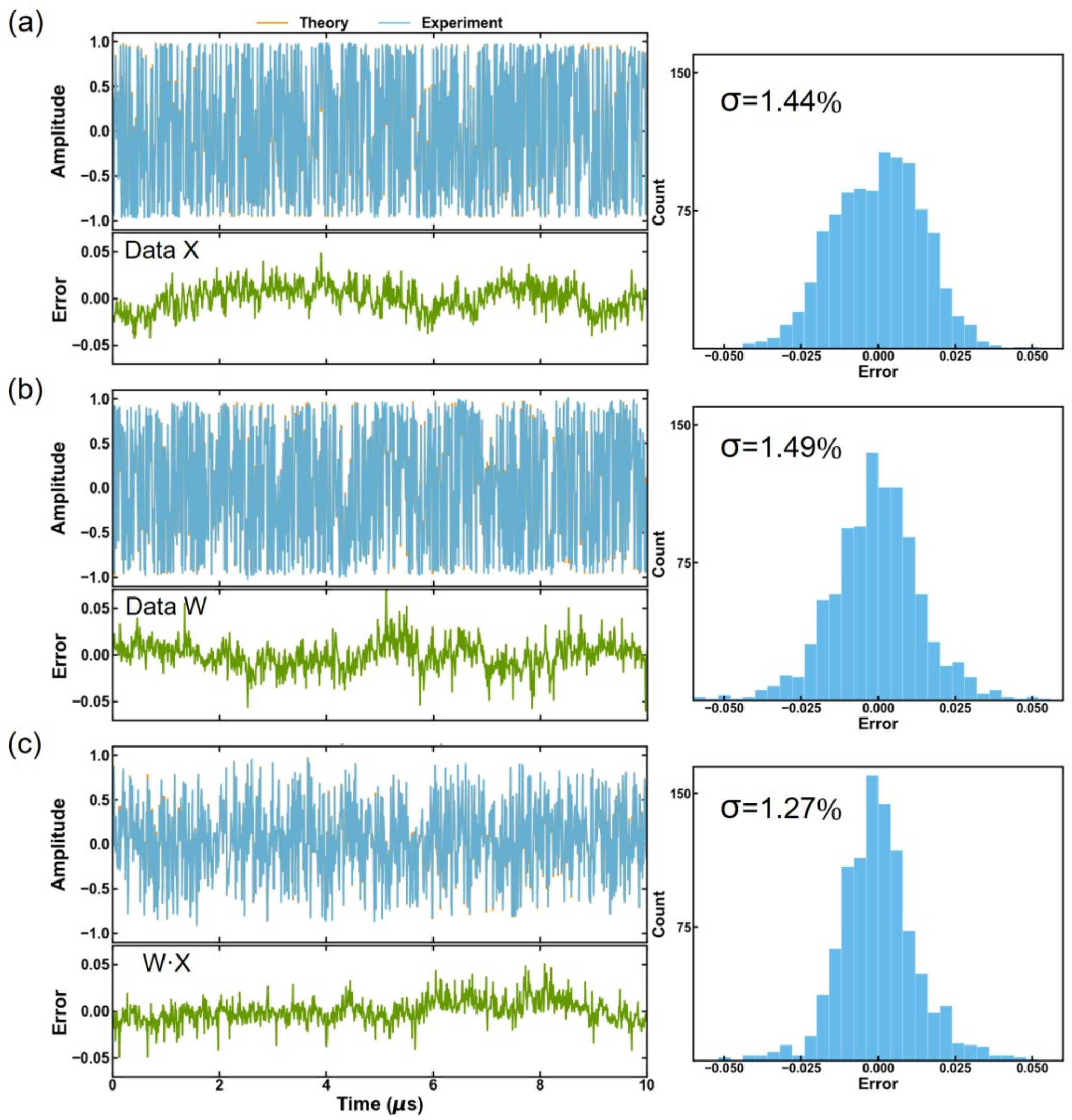

*Fig. 3. Dynamic four-quadrant optical multiplication characterization. Time-domain waveforms, absolute errors, and corresponding error histograms for: (a) bipolar input data $X_i$, (b) bipolar weight signals $W_i$, and (c) multiplication outputs $W_i \cdot X_i$.*

The experimentally measured multiplication output is presented in Fig. 3(c). Despite involving multiple physical processes, including dual-wavelength multiplexing, electro-optic modulation, wavelength selective weighting, and balanced photodetection, the measured output waveform remains highly consistent with the ideal mathematical product. The final multiplication error exhibits a low standard deviation of 1.27%, which confirms that no substantial error accumulation occurs during computation. While residual errors stem from modulator nonlinearity and non-ideal spectral responses, the dual-wavelength architecture successfully suppresses common-mode fluctuations to ensure stable signed computing fidelity within the full four-quadrant range. Significantly, the fact that both positive and negative signals can be processed without distortion across the two signal regions, with no polarity-dependent distortion, underscores the exceptional robustness of this native signed computing approach.

## Neural network classification demonstrations

To further evaluate the system level computing capability of the proposed architecture, the fabricated photonic circuit was applied to practical neural network inference tasks. We trained our chip to classify labeled noisy synthetic datasets - the moons dataset, a classic example of nonlinear classification. The dataset consists of two-dimensional vectors $X_i = [x, y]$ with an approximately zero-centered distribution, and is labeled $y_i \in \{0,1\}$, where 0 corresponds to the blue points and 1 corresponds to the red points in Fig. 4(b). To fit the complex nonlinear decision boundary, we employed a two-layer architecture, yielding an output probability in the form of:

$$\hat{y}_i = \sigma(W_2 \cdot ReLU(W_1 \cdot X_i + b_1) + b_2)$$

where the learnable parameters are contained in the weight matrices $W_1$, $W_2$ and bias vectors $b_1$, $b_2$. ReLU provides the nonlinear feature mapping for the hidden layer and $\sigma$ denotes the Sigmoid activation function that converts the output into a posterior probability. A binary cross-entropy cost function was utilized throughout the training. For each two-dimensional test vector, only 2 matrix-vector multiplications (MVMs), along with their corresponding element-wise nonlinear operations are required to compute the predicted probability $\hat{y}_i$. Based on Bayesian minimum-error-rate decision theory, a data point is classified as positive ($y_i = 1$) if and only if $\hat{y}_i \geq 0.5$; otherwise, it is classified as negative ($y_i = 0$), thereby achieving accurate classification of this nonlinear topological structure. To map this mathematical model onto the hardware, the bipolar input vectors were directly encoded using the proposed dual-wavelength signed-computing architecture and sequentially processed by the optical multiplier. The hidden-layer MVM operations were physically executed in the optical domain, while the nonlinear activation and backpropagation procedures were implemented electronically. Fig. 4(a) presents the training-accuracy evolution during the learning process. The experimentally measured photonic training trajectory closely follows the ideal electronic baseline throughout the optimization procedure, indicating accurate hardware execution of the signed linear operations. After training convergence, the photonic system achieves a classification accuracy of 95.1%, closely approaching the 95.4% accuracy of the ideal electronic model. Furthermore, comparative electronic simulations reveal that when the input data points are confined strictly to the first quadrant, both the training accuracy and convergence speed are significantly inferior to those achieved with zero-centered data. This comparative result fully underscores the necessity of developing photonic neural networks capable of executing signed operations. The corresponding decision boundary generated by the photonic hardware is shown in Fig. 4(b).

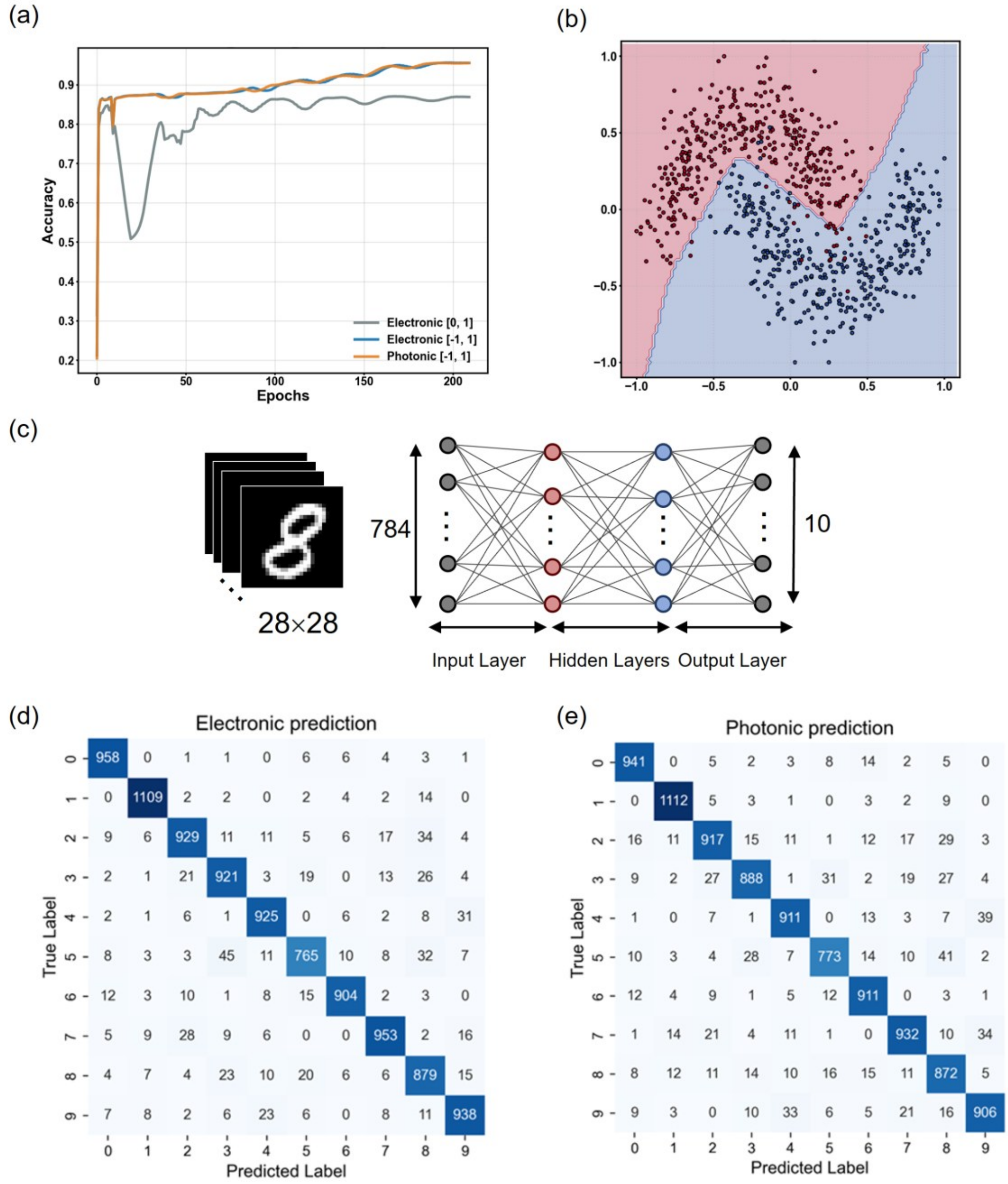

*Fig. 4. Neural network classification demonstrations. (a) Training accuracy evolution for the Two Moons task. (b) Photonic decision boundary. (c) Network topology for MNIST digit recognition. Confusion matrices for (d) the ideal electronic model and (e) the photonic hardware prediction.*

Another demonstration of our system involves using the chip to classify handwritten digits from the MNIST dataset. As shown in Fig. 4(c), we constructed a multilayer perceptron consisting of three layers: an input layer, a hidden layer, and an output layer. First, we flatten the 28 × 28 grayscale images into a vector of size n = 784, which serves as the input to the first layer. The signal then passes through a 12-dimensional hidden layer combined with a ReLU digital nonlinear activation function. Finally, the 10-dimensional output vector is processed by an electronic Softmax nonlinear activation function and a cross-entropy loss function, mapping it into ten categories representing the digits 0 to 9. Figs. 4(d) and 4(e) respectively present the confusion matrices for the ideal electronic model and the photonic predictions across 10,000 MNIST test images. Although the accumulation of analog noise and hardware imperfections cause a moderate degradation in performance, the PNN still achieves a recognition accuracy of 91.63%, closely approaching the 92.81% accuracy of the ideal electronic network.

## Discussion and outlook

The proposed dual-wavelength incoherent architecture provides a practical pathway toward scalable photonic computing systems. By encoding bipolar information directly in the spectral domain, the present approach natively executes signed computation within a shared physical path. This intrinsic configuration effectively circumvents inter-path mismatch and significantly reduces calibration complexity. It is worth noting that while scaling the network to accommodate higher-dimensional matrix operations inherently increases the overall physical footprint, our spectral multiplexing strategy ensures that the hardware overhead required for signed processing remains strictly constant for each individual computing channel.

Building on this advantage, the architecture can be extended through a hybrid scaling strategy that combines spectral domain multiplexing and spatial domain fan-out. At the input stage, multiple wavelength pairs can be introduced, where each pair encodes complementary signal components with a designed phase relation, enabling parallel signed-computing channels within a shared physical path. On the weight side, optical power splitting networks distribute signals into multiple computing units, each programmed with independent weights, thereby enabling large-scale parallel multiply–accumulate operations with high hardware efficiency. Further performance improvements can be achieved through refined spectral engineering. Optimizing the spectral responses of the weighting MZI and wavelength allocation strategies can enhance complementary accuracy and reduce residual imbalance between wavelength channels. Meanwhile, increasing the number of wavelength pairs enables higher degrees of parallelism without fundamentally altering the system architecture. From a system perspective, incorporating on-chip nonlinear activation functions will be essential for realizing fully functional photonic neural networks. Such nonlinear operations may be implemented using tunable high-Q resonant structures[34,35] or the periodic poling technique.[36,37] In addition, integration with high-speed electronic drivers and photodetectors is expected to fully exploit the ultrabroadband electro-optic capability of the thin-film lithium niobate platform.[38] These characteristics collectively point toward scalable, high-throughput, and robust photonic computing architectures for future large-scale information processing systems.

## Conclusion

In conclusion, we experimentally demonstrate a native signed incoherent optical computing architecture on a thin-film lithium niobate platform. By employing dual-wavelength differential encoding within a shared physical structure, the proposed scheme enables four-quadrant optical multiplication without requiring coherent phase stabilization or duplicated weighting paths. The demonstrated system achieves high computational fidelity and broadband operation with a 3 dB bandwidth exceeding 40 GHz. More importantly, the proposed signed-computing paradigm eliminates the additional hardware scaling overhead. The combination of native signed computation, wavelength-pair-based spectral encoding, and constant-overhead scalability establishes a promising hardware foundation for next-generation high-throughput photonic computing systems.

## Research funding

This research was funded by National Key R&D Program of China (2025YFF0524600). National Natural Science Foundation of China (12192251, 12334014, 62335019, 12134001, 12304418, 12474378). Quantum Science and Technology-National Science and Technology Major Project (2021ZD0301403). Shanghai Municipal Science and Technology Major Project (Grant No.2019SHZDZX01)

## Author contributions

All authors have accepted responsibility for the entire content of this manuscript and approved its submission.

## Conflict of interest

The authors declare no conflict of interest.

## Data availability

The datasets generated and/or analyzed during the current study are available from the corresponding author upon reasonable request.